\documentclass[11pt]{article}
\usepackage{graphics,latexsym,amsfonts}
\usepackage{graphicx}
\usepackage{float}
\usepackage{picture}
\usepackage[american]{babel}
\usepackage{amsmath,mathtools,amsthm}
\usepackage{enumitem}
\usepackage{color}
\usepackage{comment}
\usepackage{lineno}
\usepackage{algorithm}
\usepackage{algpseudocode}
\usepackage{caption}

\author{
A.\ Pluchino\thanks{Department\ of Physics and Astronomy, University\ of Catania and INFN Sezione di Catania, Italy; alessandro.pluchino@ct.infn.it}$\:$,
A.\ E.\ Biondo\thanks{Dept.\ of Economics and Business, Univ.\ of Catania, Italy; ae.biondo@unict.it}$\:$,
A.\ Rapisarda\thanks{Department\ of Physics and Astronomy, University\ of Catania and INFN Sezione di Catania, Italy; Complexity Science Hub Vienna; andrea.rapisarda@ct.infn.it}}

\title{\bf Talent vs Luck:\\ the role of randomness in success and failure} 

\begin{document}
\date{ }
\maketitle

\begin{abstract}
The largely dominant  meritocratic paradigm of highly competitive Western cultures is rooted on the belief that success is due mainly, if not exclusively, to personal qualities such as  talent, intelligence, skills, smartness, efforts, willfulness, hard work or risk taking. Sometimes, we are willing to admit that a certain degree of luck could also play a role in achieving significant material success. But, as a matter of fact, it is rather common to  underestimate the importance of external forces in individual successful stories. It is very well known that intelligence (or, more in general, \textit{talent} and personal qualities) exhibits a Gaussian distribution among the population, whereas the distribution of wealth - often considered a proxy of success - follows typically a power law (Pareto law), with a large majority of poor people and a very small number of billionaires. Such a discrepancy between a Normal distribution of inputs, with a typical scale (the average talent or intelligence), and the scale invariant distribution of outputs,  suggests that some hidden ingredient is at work behind the scenes. In this paper, with the help of a very simple agent-based toy model, we suggest that such an ingredient is just randomness. In particular, we show that, if it is true that some degree of talent is  necessary to be successful in life, almost never the most talented people reach the highest peaks of success, being overtaken by mediocre but sensibly luckier individuals. As to our knowledge, this counterintuitive result - although implicitly suggested between the lines in a vast literature - is quantified here for the first time. It sheds new light on the effectiveness of assessing merit on the basis of the reached level of success and underlines the risks of distributing excessive honors or resources to people who, at the end of the day, could have been simply luckier than others. With the help of this model, several policy hypotheses are also addressed and compared  to show the most efficient strategies  for public funding of research in order to improve meritocracy, diversity and innovation.

\medskip
\noindent \textbf{Keywords:} Success, Talent, Luck, Randomness, Serendipity, Funding strategies.
\end{abstract}

\section{Introduction}
The ubiquity of power-law distributions in many physical, biological or socio-economical complex systems can be seen as a sort of mathematical signature of  their strongly correlated dynamic behavior and  their scale invariant topological structure \cite{SOC,Barabasi,Newman,Tsallis}. In socio-economic context, after Pareto's work \cite{Pareto,Steindl,Atkinson,Persky,Klass}, it is well known that the wealth distribution follows a power-law, whose typical long tailed shape reflects the deep existing gap between the rich and the poor in our society. A very recent report \cite{Oxfam} shows that today this gap is far greater than it had been feared: eight men own the same wealth as the 3.6 billion people constituting the poorest half of humanity. In the last 20 years, several theoretical models have been developed to derive the wealth distribution in the context of statistical physics and probability theory, often adopting a multi-agent perspective with a simple underlying dynamics \cite{Bouchaud,Dragulescu,Chakraborti,Patriarca,Scalas1,Scalas2,Scalas3}. 

Moving along this line, if one considers the individual wealth as a proxy of success, one could argue that its deeply asymmetric and unequal distribution among people is either a consequence of their natural differences in talent, skill, competence, intelligence, ability or a measure of their willfulness, hard work or determination. Such an assumption is, indirectly, at the basis of the so-called {\it meritocratic paradigm}: it affects not only the way our society grants work opportunities, fame and honors, but also the strategies adopted by Governments in assigning resources and funds to those who are considered the most deserving individuals.

However, the previous conclusion appears to be in strict contrast with the accepted evidence that human features and qualities cited above are normally distributed among the population, i.e. follow a symmetric Gaussian distribution around a given mean. For example, intelligence, as measured by IQ tests, follows this pattern: average IQ is 100, but nobody has an IQ of 1,000 or 10,000. The same holds for efforts, as measured by hours worked: someone works more hours than the average and someone less, but nobody works a billion times more hours than anybody else. 

On the other hand, there is nowadays an ever greater evidence about the fundamental role of chance, luck or, more in general, random factors, in determining successes or failures in our personal and professional lives. In particular, it has been shown that scientists have the same chance along their career of publishing their biggest hit \cite{Sinatra}; that those with earlier surname initials are significantly more likely to receive tenure at top departments \cite{Einav}; that the distributions of bibliometric indicators collected by a scholar might be the result of chance and noise  related to multiplicative phenomena connected to a {\it publish or perish} inflationary mechanism \cite{Ruocco}; that one's position in an alphabetically sorted list may be important in determining access to over-subscribed public services \cite{Jurajda}; that middle name initials enhance evaluations of intellectual performance \cite{Tilburg}; that people with easy-to-pronounce names are judged more positively than those with difficult-to-pronounce names \cite{Laham}; that individuals with noble-sounding surnames are found to work more often as managers than as employees \cite{Silberzahn}; that females with masculine monikers are more successful in legal careers \cite{Coffey}; that roughly half of the variance in incomes across persons worldwide is explained only by their country of residence and by the income distribution within that country \cite{Milanovic}; that the probability of becoming a CEO is strongly influenced by your name or by your month of birth \cite{Du,Deaner,Brooks}; that the innovative ideas are the results of a random walk in our brain network \cite{Iacopini}; and that even the probability of developing a cancer, maybe cutting a brilliant career, is mainly due to simple bad luck \cite{Tomasetti,Newgreen}. Recent studies on lifetime reproductive success further corroborate these statements showing that, if trait variation may influence the fate of populations, luck often governs the lives of individuals \cite{Snyder1,Snyder2}. 
   
In recent years many authors, among whom the statistician and risk analyst Nassim N. Taleb \cite{Taleb1,Taleb2}, the investment strategist Michael Mauboussin \cite{Mauboussin} and the economist Robert H. Frank \cite{Frank}, have explored in several successful books the relationship between luck and skill in financial trading, business, sports, art, music, literature, science and in many other fields. They reach the conclusion that chance events play a much larger role in life than many people once imagined. Actually, they do not suggest that success is independent of talent and efforts, since in highly competitive arenas or 'winner-takes-all' markets, like those where we live and work today, people performing well are almost always extremely talented and hard-working. Simply, they conclude that talent and efforts are not enough: you have to be also in the right place at the right time. In short: luck also matters, even if its role is almost always underestimated by successful people. This happens because randomness often plays out in subtle ways, therefore it is easy to construct narratives that portray success as having been inevitable. Taleb calls this tendency "narrative fallacy" \cite{Taleb2}, while the sociologist Paul Lazarsfeld adopts the terminology "hindsight bias". In his recent book "Everything Is Obvious: Once You Know the Answer" \cite{Watts}, the sociologist and network science pioneer Duncan J. Watts, suggests that both narrative fallacy and hindsight bias operate with particular force when people observe unusually successful outcomes and consider them as the necessary product of hard work and talent, while they mainly emerge from a complex and interwoven sequence of steps, each depending on precedent ones: if any of them had been different, an entire career or life trajectory would almost surely differ too. This argument is also based on the results of a seminal experimental study, performed some years before by Watts himself in collaboration with other authors \cite{Salganik}, where the success of previously unknown songs in an artificial music market was shown not to be correlated with the quality of the song itself. And this clearly makes very difficult any kind of prediction, as also shown in another more recent study \cite{Travis}.   

In this paper, by adopting an agent-based statistical approach, we try to realistically quantify the role of luck and talent in successful careers. In section 2, building on a minimal number of assumptions, i.e. a Gaussian distribution of talent \cite{Stewart} and a multiplicative dynamics for both successes and failures \cite{Sinha}, we present a simple model, that we call  "Talent vs Luck" (TvL) model, which mimics the evolution of careers of a group of people over a working period of 40 years. The model shows that, actually, randomness plays a fundamental role in selecting the most successful individuals. It is true that, as one could expect, talented people are more likely to become rich, famous or important during their life with respect to poorly equipped ones. But - and this is a less intuitive rationale - ordinary people with an average level of talent are statistically destined to be successful (i.e. to be placed along the tail of some power law distribution of success) much more than the most talented ones, provided that they are more blessed by fortune along their life. This fact is commonly experienced, as pointed in refs.\cite{Taleb1,Taleb2,Frank}, but, to our knowledge, it is modeled and quantified here for the first time.

The success of the averagely-talented people strongly challenges the "meritocratic" paradigm and all those strategies and mechanisms, which give more rewards, opportunities, honors, fame and resources to people considered the best in their field \cite{Fortin,Jacob}. The point is that, in the vast majority of cases, all evaluations of someone's talent are carried out {\it a posteriori}, just by looking at his/her performances - or at reached results -  in some specific area of our society like sport, business, finance, art, science, etc. This kind of misleading evaluation ends up switching cause and effect, rating as the most talented people those who are, simply, the luckiest ones \cite{Aguinis,Denrell}. In line with this perspective, in previous works, it was advanced a warning against such a kind of "naive meritocracy" and it was shown the effectiveness of alternative strategies based on random choices in many different contexts, such as management, politics and finance \cite{Pluchino1,Pluchino2,Pluchino3,Biondo1,Biondo2,Biondo3,Biondo4,Biondo5}. In section 3 we provide an application of our approach and sketch a comparison of possible public funds attribution schemes in the scientific research context. We study the effects of several distributive strategies, among which the "naively" meritocratic one, with the aim of exploring new ways to increase both the minimum level of success of the most talented people in a community and the resulting efficiency of the public expenditure. We also explore, in general, how opportunities offered by the environment, as the education and income levels (i.e., external factors depending on the country and the social context where individuals come from), do matter in increasing probability of success. Final conclusive remarks close the paper.

 \section{The Model}

In what follows we propose an agent-based model, called "Talent vs Luck" (TvL)  model, which builds on a small set of very simple assumptions, aiming to describe the evolution of careers of a group of people influenced by lucky or unlucky random events. 

\begin{figure}
\begin{center}
\includegraphics[width=3.5 in,angle=0]{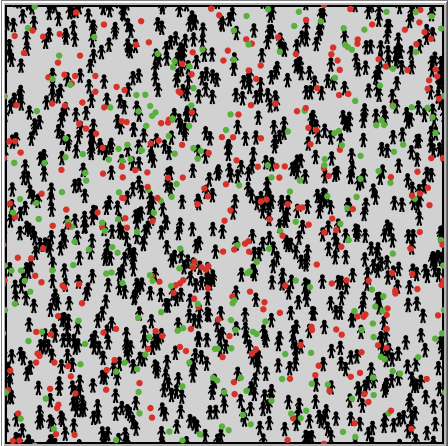}
\caption{\small 
An example of initial setup for our simulations. All the simulations presented in this paper were  realized within the NetLogo agent-based model environment \cite{Netlogo}. $N=1000$ individuals (agents), with different degrees of talent (intelligence, skills, etc.), are randomly located in their fixed positions within a square world of $201x201$ patches with periodic boundary conditions. During each simulation, which covers several dozens of years, they are exposed to a certain number $N_E$ of lucky (green circles) and unlucky (red circles) events, which move across the world following random trajectories (random walks). In this example $N_E=500$. 
}
\label{world} 
\end{center}
\end{figure}

We consider $N$ individuals, with talent $T_i$ (intelligence, skills, ability, etc.) normally distributed in the interval $[0,1]$ around a given mean $m_T$ with a standard deviation $\sigma_T$, randomly placed in fixed positions within a square world (see Figure \ref{world}) with periodic boundary conditions (i.e. with a toroidal topology) and surrounded by a certain number $N_E$ of "moving"  events (indicated by dots), someone lucky, someone else unlucky (neutral events are not considered in the model, since they have not relevant effects on the individual life). In Figure \ref{world} we report these events as colored points: lucky ones, in green and with relative percentage $p_L$, and unlucky ones, in red and with percentage ($100 - p_L$). The total number of event-points $N_E$ are uniformly distributed, but of course such a distribution would be perfectly uniform only for $N_E \rightarrow \infty$. In our simulations, typically will be $N_E \sim N/2$: thus, at the beginning of each simulation, there will be a greater random concentration of lucky or unlucky event-points in different areas of the world, while other areas will be more neutral. The further random movement of the points inside the square lattice, the world, does not change this fundamental features of the model, which exposes different individuals to different amount of lucky or unlucky events during their life, regardless of their own talent.   

\begin{figure}[t]
\begin{center}
\includegraphics[width=4.0in,angle=0]{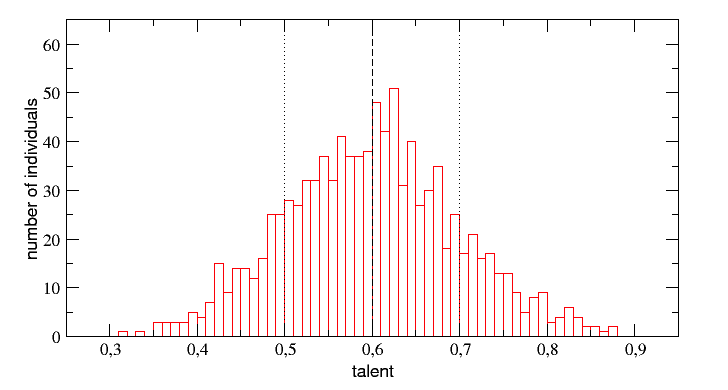}
\caption{\small 
Normal distribution of talent among the the population (with mean $m_T=0.6$, indicated by a dashed vertical line, and standard deviation $\sigma_T=0.1$ - the values $m_T \pm \sigma_T$ are indicated by two dotted vertical lines). This distribution is truncated in the interval $[0,1]$ and does not change during the simulation.        
}
\label{talent} 
\end{center}
\end{figure}

For a single simulation run, a working life period $P$ of $40$ years (from the age of twenty to the age of sixty) is considered, with a time step $\delta_t$ equal to six months. At the beginning of the simulation, all agents are endowed with the same amount $C_i=C(0)$ $\forall i=1,...,N$ of capital, representing their starting level of success/wealth. This choice has the evident purpose of not offering any initial advantage to anyone. While the agents' talent is time-independent, agents' capital changes in time. During the time evolution of the model, i.e. during the considered agents' life period, all event-points move randomly  around the world and, in doing so, they possibly intersect the position of some agent. More in detail, at each time each event-point covers a distance of $2$ patches in a random direction. We say that an intersection does occur for an individual when an event-point is present inside a circle of radius $1$ patch centered on the agent (the event-point does not disappear after the intersection). Depending on such an occurrence, at a given time step $t$ (i.e. every six months), there are three different possible actions for a given agent $A_k$:

\begin{enumerate}

\item No event-point intercepts the position of agent $A_k$: this means that no relevant facts have happened during the last six months; agent $A_k$ does not perform any action.
\item A lucky event intercepts the position of agent $A_k$: this means that a lucky event has occurred during the last six month (notice that, in line with ref.\cite{Iacopini}, also the production of an innovative idea is here considered as a lucky event occurring in the agent's brain); as a consequence, agent $A_k$ doubles her capital/success with a probability proportional to her talent $T_k$. It will be $C_k(t)=2C_k(t-1)$ only if $rand[0,1] < T_k$, i.e. if the agent is smart enough to profit from his/her luck. 
\item An unlucky event intercepts the position of agent $A_k$: this means that an unlucky event has occurred during the last six month; as a consequence, agent $A_k$ halves her capital/success, i.e. $C_k(t)=C_k(t-1)/2$.                      
\end{enumerate}

The previous agents' rules (including the choice of dividing by a factor of 2 the initial capital in case of unlucky events and doubling it in case of lucky ones, proportionally to the agent's talent), are intentionally simple and can be considered widely shareable, since they are based on the common sense evidence that success, in everyone life, has the property to both grow or decrease very rapidly. Furthermore, these rules gives a significant advantage to highly talented people, since they can make much better use of the opportunities offered by luck (including the ability to exploit   a good idea born in their brains). On the other hand, a car accident or a sudden desease, for example, are always unlucky events where talent plays no role. In this respect, we could more effectively generalise the definition of "talent" by identifying it with "any personal quality which enhances the chance to grab an opportunity". In other words, by the term "talent" we broadly mean intelligence, skill, smartness, stubbornness, determination, hard work, risk taking and so on. What we will see in the following is that the advantage of having a great talent is a {\it necessary, but not a sufficient}, condition to reach a very high degree of success.  

\begin{figure}[t]
\begin{center}
\includegraphics[width=5.0in,angle=0]{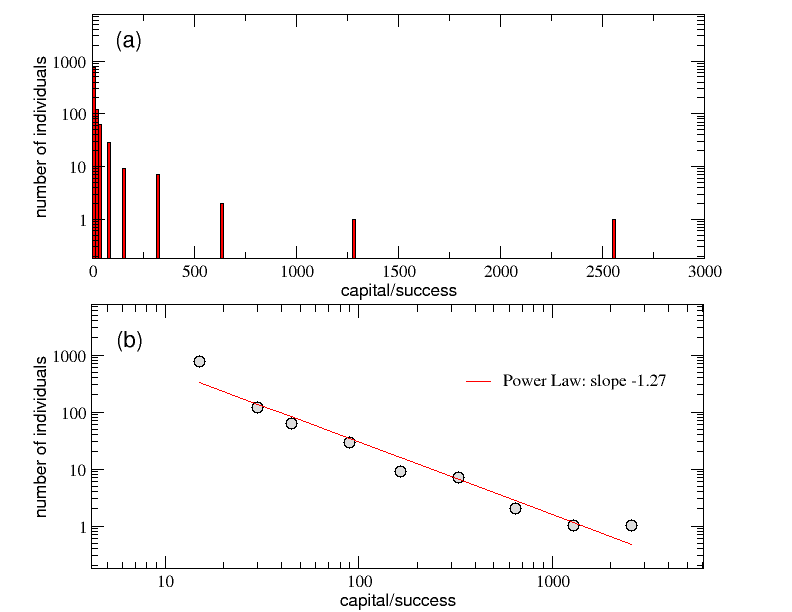}
\caption{\small 
Final distribution of capital/success among the population, both in log-lin (a) and in log-log (b) scale. Despite the normal distribution of talent, the tail of distribution of success - as visible in panel (b) - can be well fitted with a power-law curve with slope $-1.27$. We also verified that the capital/success distribution follows the Pareto's "80-20" rule, since $20\%$ of the population owns  $80\%$ of the total capital, while the remaining $80\%$ owns the $20\%$ of the capital. 
}
\label{capital-pdf} 
\end{center}
\end{figure}

\subsection{Single run results}

In this subsection we present the results of a typical single run simulation. Actually, such results are very robust so, as we will show later, they can be considered largely representative of the general framework emerging from our model. 

Let us consider $N=1000$ agents, with a starting equal amount of capital $C(0)=10$ (in dimensionless units) and with a fixed talent $T_i \in[0,1]$, which follows a normal distribution with mean $m_T=0.6$ and standard deviation $\sigma_T=0.1$ (see Figure \ref{talent}). As previously written, the simulation spans a realistic time period of $P=40$ years, evolving through time steps of six months each, for a total of $I=80$ iterations. In this simulation we consider $N_E=500$ event-points, with a percentage $p_L=50\%$ of lucky events. 

 At the end of the simulation, as shown in panel (a) of Figure \ref{capital-pdf}, we find that the simple dynamical rules of the model are able to produce an unequal distribution of capital/success, with a large amount of  poor (unsuccessful) agents  and a small number of very rich (successful) ones. Plotting the same distribution in log-log scale in panel (b) of the same Figure, a Pareto-like power-law distribution is observed, whose tail is  well fitted by the function $y(C) \sim C^{-1.27}$. Therefore, despite the normal distribution of talent, the TvL model seems able to capture the first important feature observed in  the comparison with real data: the deep existing gap between rich and poor and its scale invariant nature. In particular, in our simulation, only 4 individuals have more than $500$ units of capital and the 20 most successful individuals hold the $44\%$ of the total amount of capital, while almost half of the population stay under $10$ units. Globally, the Pareto's "80-20" rule is respected, since the $80\%$ of the population owns only the $20\%$ of the total capital, while the remaining $20\%$ owns the $80\%$ of the same capital. Although this disparity surely seems unfair, it would be to some extent acceptable if the most successful people were the most talented one, so deserving to have accumulated more capital/success with respect to the others. But are things really like that? 

\begin{figure}[t]
\begin{center}
\includegraphics[width=5.0in,angle=0]{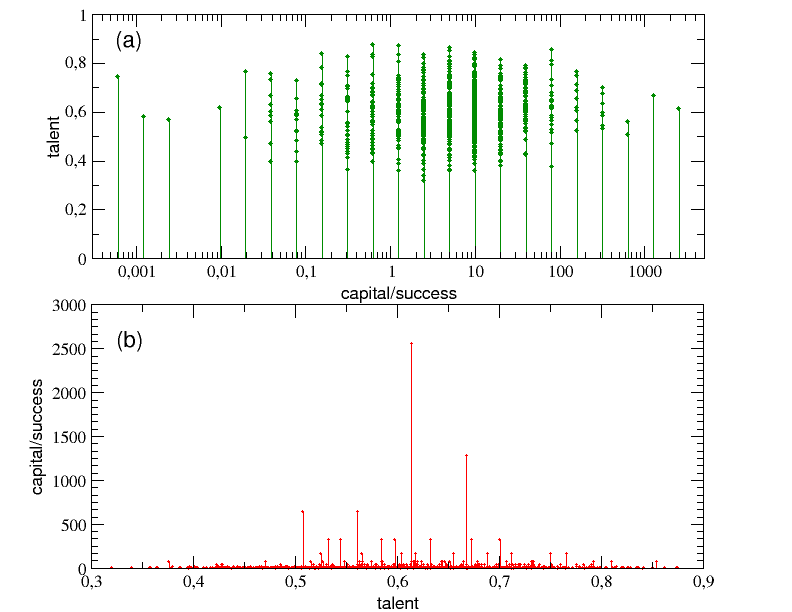}
\caption{\small 
In panel (a) talent is plotted as function of capital/success (in logarithmic scale for a better visualization): it is evident  that the most successful individuals are not the most talented ones. In panel (b), vice-versa, capital/success is plotted as function of talent: here, it can be further appreciated the fact that the most successful agent, with $C_{max}=2560$, has a talent only slightly greater than the mean value $m_T=0.6$, while the most talented one has a capital/success lower than $C=1$ unit, much less of the initial capital $C(0)$. See text for further details.
}
\label{talent-success} 
\end{center}
\end{figure}

In panels (a) and (b) of Figure \ref{talent-success}, respectively, talent is plotted as function of the final capital/success and vice-versa (notice that, in panel (a), the capital/success takes only discontinuous values: this is due to the choice of having used an integer initial capital equal for all the agents). Looking at both panels, it is evident  that, on one hand, the most successful individuals are not the most talented ones and, on the other hand, the most talented individuals are not the most successful ones. In particular, the most successful individual, with $C_{max}=2560$, has a talent $T^*=0.61$, only slightly greater than the mean value $m_T=0.6$, while the most talented one ($T_{max}=0.89$) has a capital/success lower than $1$ unit ($C=0.625$). 

As we will see more in detail  in the next subsection, such a result  is not a special case, but it is rather the rule for this kind of system: the maximum success never coincides with the maximum talent, and vice-versa. Moreover, such a misalignment between success and talent is disproportionate and highly nonlinear. In fact, the average capital of all people with talent $T>T^*$ is $C\sim20$: in other words, the capital/success of the most successful individual, who is moderately gifted, is $128$ times greater than the average capital/success of people who are more talented  than him. We can conclude that, if there is not an exceptional talent behind the enormous success of some people, another factor is probably at work. Our simulation clearly shows that such a factor is just pure luck. 

\begin{figure}[t]
\begin{center}
\includegraphics[width=2.95in,angle=0]{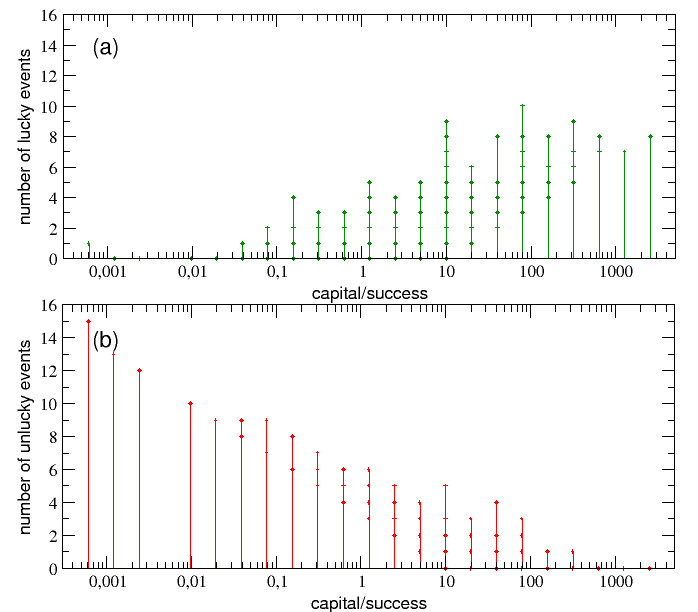}
\includegraphics[width=3.2in,angle=0]{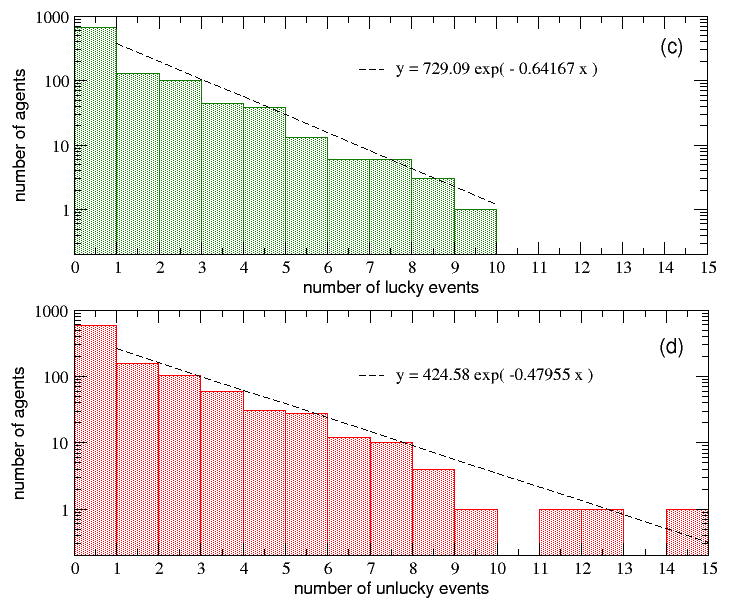}
\caption{\small 
Total number of lucky events (a) or unlucky events (b)  as function of the capital/success of the agents. The plot shows  the existence of a strong correlation between success and luck: the most successful individuals are also the luckiest ones, while the less successful are also the unluckiest ones. Again, having used an initial capital equal for all the agents, it follows  that several events are grouped in discontinuous values of the capital/success. In panels (c) and (d) the frequency distributions of, respectively, the number of lucky and unlucky events are reported in log-linear scale. As visible, both the distributions can be well fitted by exponential distributions with similar negative exponents.}
\label{luck-unluck} 
\end{center}
\end{figure}

In Figure \ref{luck-unluck} the number of lucky and unlucky events occurred to all people during their working lives is reported as a function of their final capital/success. Looking at panel (a), it is evident that the most successful individuals are also the luckiest ones (notice that it in this panel are reported all the lucky events occurred to the agents and not just those that they took advantage of, proportionally to their talent). On the contrary, looking at panel (b), it results that the less successful individuals are also the unluckiest ones. In other words, although there is an absence of correlation between success and talent coming out of the simulations, there is also a  very strong correlation between success and luck. Analyzing the details of the frequency distributions of the number of lucky or unlucky events occurred to individuals, we found - as shown in panels (c) and (d) - that both of them are exponential, with exponents $0.64$ and $0.48$, and averages $1.35$ and $1.66$, respectively, and that the maximum numbers of lucky or unlucky events occurred were, respectively, $10$ and $15$. Moreover about $16\%$ of people had a "neutral" life, without lucky or unlucky events at all, while about $40\%$ of individuals exclusively experienced only one type of events (lucky or unlucky).   

\begin{figure}[t]
\begin{center}
\includegraphics[width=5.5in,angle=0]{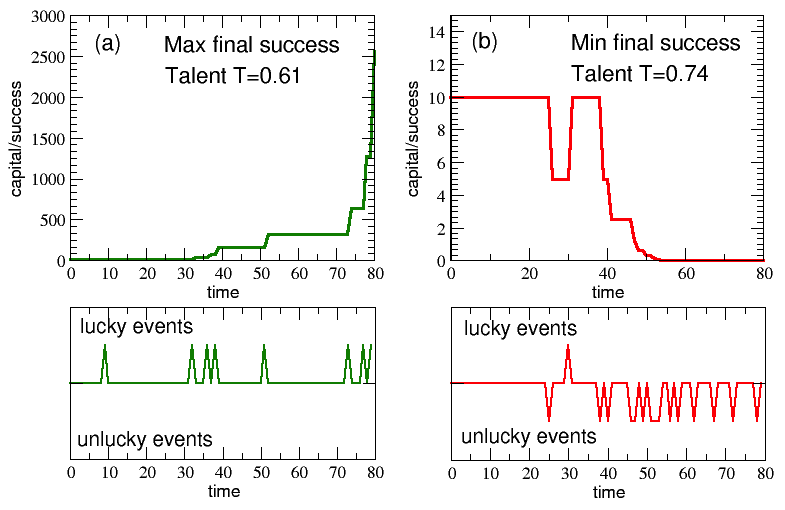}
\caption{\small 
(a) Time evolution of success/capital for the most successful individual and (b) for the less successful one, compared with the corresponding sequences of lucky or unlucky events occurred during their working lives ($80$ semesters, i.e. $40$ years). The time occurrence of these events is indicated, in the bottom panels, with upwards or downwards spikes.    
}
\label{success-evolution} 
\end{center}
\end{figure}

It is also interesting to look at the time evolution of the success/capital of both the most successful individual and the less successful one, compared with the corresponding sequence of lucky or unlucky events occurred during the $40$ years ($80$ time steps, one every 6 months) of their working  life. This can be observed, respectively, in the left and the right part of Figure \ref{success-evolution}. Differently from the panel (a) of Figure \ref{luck-unluck}, in the bottom panels of this figure only lucky events that agents have taken advantage of thanks to their talent, are shown.  

In panels (a), concerning the moderately talented but most successful individual, it clearly appears that, after about a first half of his working  life with a low occurrence of lucky events (bottom panel), and then with a low level of capital (top panel), a sudden concentration of favorable events between $30$ and $40$ time steps (i.e. just before the age of $40$ of the agent) produces a rapid increase in capital, which becomes exponential in the last $10$ time steps (i.e. the last 5 years of the agent's career), going from $C=320$ to $C_{max}=2560$. 

On the other hand, looking at (top and bottom) panels (b), concerning the less successful individual, it is evident that a particularly unlucky second half of his working  life, with a dozen of unfavorable events, progressively reduces the capital/success bringing it at its final value of $C=0.00061$. It is interesting to notice that this poor agent had, however, a talent $T=0.74$ which was greater than that of the most successful agent. Clearly, good luck made the difference. And, if it is true that the most successful agent has had the merit of taking advantage of all the opportunities presented to him (in spite of his average talent), it is also true that if your life is as unlucky and poor of opportunities as that of the other agent, even a great talent becomes useless against the fury of misfortune.              

All the results shown in this subsection for a single simulation run\footnote{A demo version of the NetLogo code of the TvL model used  for the single run simulations can be found on the Open ABM repository - https://www.comses.net/codebases/} are very robust and, as we will see in the next subsection, they persist, with small differences, if we repeat many times the simulations starting with the same talent distribution, but with a different random positions of the individuals.

\subsection{Multiple runs results}

In this subsection we present the global results of a simulation averaging over  $100$ runs, each starting with different random initial conditions. The values of the control parameters are the same of those used  in the previous subsection: $N=1000$ individuals, $m_T=0.6$ and  $\sigma_T=0.1$ for the normal talent distribution, $I=80$ iteration (each one representing $\delta_t=6$ months of working  life), $C(0)=10$ units of initial capital, $N_E=500$ event-points and a percentage $p_L=50\%$ of lucky events.

\begin{figure}
\begin{center}
\includegraphics[width=5.0in,angle=0]{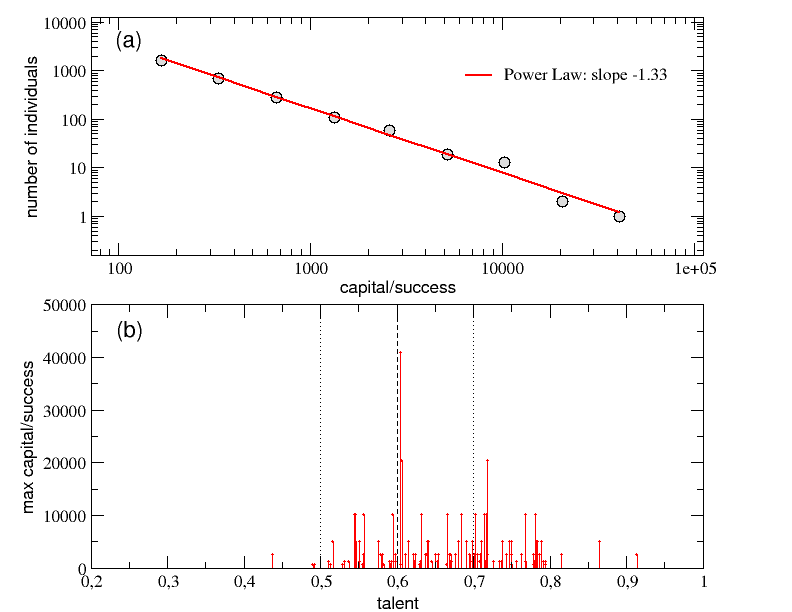}
\caption{\small 
Panel (a): Distribution of the final capital/success calculated over $100$ runs for a population with different random initial conditions. The distribution can be well fitted with a power-law curve with a slope $-1.33$. 
Panel (b): The final capital $C_{max}$ of the most successful individual in each of the $100$ runs is reported as function of their talent. People with a medium-high talent result to be, on average, more successful than people with low or medium-low talent, but very often the most successful individual is a moderately gifted agent and only rarely the most talented one. The $m_T$ value, together with the values $m_T \pm \sigma_T$, are also reported as vertical dashed and dot lines respectively. 
}
\label{capital-multiple} 
\end{center}
\end{figure}

In panel (a) of Figure \ref{capital-multiple}, the global distribution of the final capital/success for all the agents collected over the $100$ runs is shown in log-log scale and it is well fitted by a power law curve with slope $-1.33$. The scale invariant behavior of capital and the consequent strong inequality among individuals, together with the Pareto's "80-20" rule observed in the single run simulation, are therefore conserved also in the case of multiple runs. Indeed, the gap between rich (successful) and poor (unsuccessful) agents has even increased, since the capital of the most successful people surpass now the $40000$ units. 

This last result can be better appreciated looking at panel (b), where the final capital $C_{max}$ of the most successful individuals only, i.e. of the best performers for each one of the $100$ runs, is reported as function of their talent. The best score was realized by an agent with a talent $T_{best}=0.6048$, practically coinciding with the mean of the talent distribution ($m_T=0.6$), who reached a peak of capital $C_{best}=40960$. On the other hand, the most talented among the most successful individuals, with a talent $T_{max}=0.91$, accumulated a capital $C_{max}=2560$, equal to only $6\%$ of  $C_{best}$.

\begin{figure}
\begin{center}
\includegraphics[width=3.1in,angle=0]{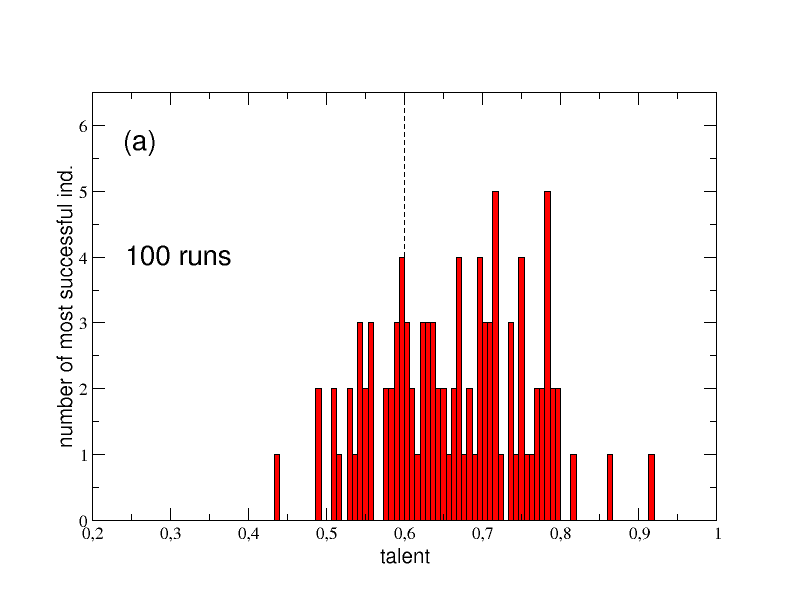}
\includegraphics[width=3.1in,angle=0]{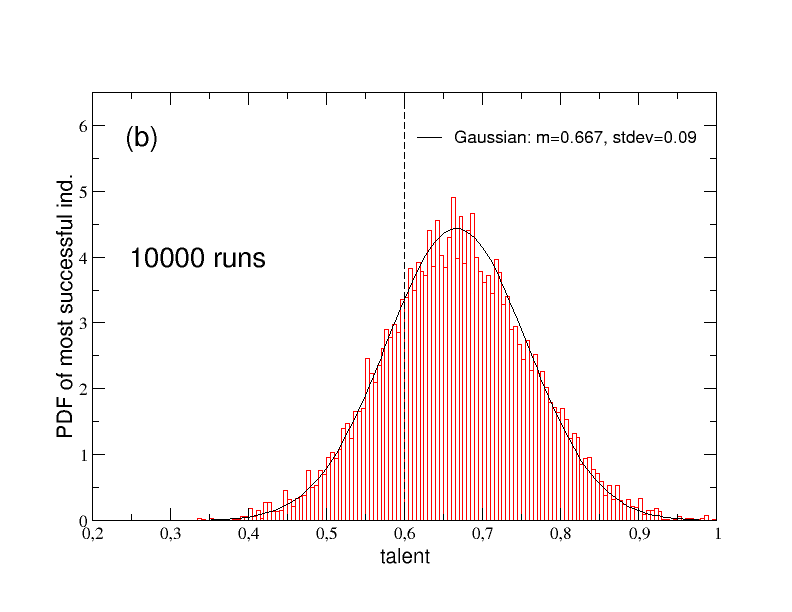}
\caption{\small 
(a) Talent distribution of the most successful individuals (best performers) in each of the $100$ runs. (b) Probability distribution function of talent of the most successful individuals calculated over $10000$ runs: it is well fitted by a normal distribution with mean $0.667$ and standard deviation $0.09$ (solid line). The mean $m_T=0.6$ of the original normal distribution of talent in the population is reported for comparison as a vertical dashed line in both  panels. 
}
\label{talent-max-success} 
\end{center}
\end{figure}

To address this point in more detail, in Figure \ref{talent-max-success} (a) we plot the talent distribution of the best performers calculated over  $100$ runs. The distribution seems to be shifted to the right of the talent axis, with a mean value $T_{av}=0.66>m_T$: this confirms, on one hand, that a medium-high talent is often necessary to reach a great success; but, on the other hand, it also indicates that it is almost never sufficient, since agents with the highest talent (e.g. with $T>m_T+2\sigma_T$, i.e. with $T>0.8$) result to be the best performers only in  $3\%$ of cases, and their capital/success never exceeds the $13\%$ of  $C_{best}$. 

In Figure \ref{talent-max-success} (b) the same distribution (normalized to unitary area in order to obtain a PDF) is calculated over $10000$ runs, in order to appreciate its true shape: it appears to be well fitted by a Gaussian $G(T)$ with average $T_{av}=0.667$ and standard deviation $0.09$ (solid line). This definitely confirms that the talent distribution of the best performers is shifted to the right of the talent axis with respect to the original distribution of talent. More precisely, this means that the conditional probability $P(C_{max}|T) = G(T) dT$ to find among the best performers an individual with talent in the interval $[T,T+dT]$ increases with the talent $T$, reaches a maximum around a medium-high talent $T_{av}=0.66$, then rapidly decreases for higher values of talent. In other words, the probability to find a moderately talented individual at the top of success is higher than that of finding there a very talented one. Notice that, in a ideal world in which talent were the main cause of success, one expects  $P(C_{max}|T)$ to be an increasing function of T. Therefore, we can conclude that the observed Gaussian shape of $P(C_{max}|T)$ is the proof that luck matters more than talent in reaching very high levels of success.      

It is also interesting to compare the average capital/success $C_{mt}\sim63$, over $100$ runs, of the most talented people and the corresponding average capital/success $C_{at}\sim33$ of people with talent very close to the mean $m_T$. We found in both cases quite small values (although greater than the initial capital $C(0)=10$), but the fact that $C_{mt}>C_{at}$ indicates that, even if the probability to find a moderately talented individual at the top of success is higher than that of finding there a very talented one, the most talented individuals of each run have, on average, more success than moderately gifted people. On the other hand, looking at the average percentage, over the $100$ runs, of individuals with talent $T > 0.7$ (i.e. greater than one standard deviation from the average) and with a final success/capital $C_{end}>10$, calculated with respect to all the agents with talent $T > 0.7$ (who are, on average for each run, $\sim 160$), we found that this percentage is equal to $32\%$: this means that the aggregate performance of the most talented people in our population remains, on average, relatively small since only one third of them reaches a final capital greater than the initial one. 

\begin{figure}[t]
\begin{center}
\includegraphics[width=4.5in,angle=0]{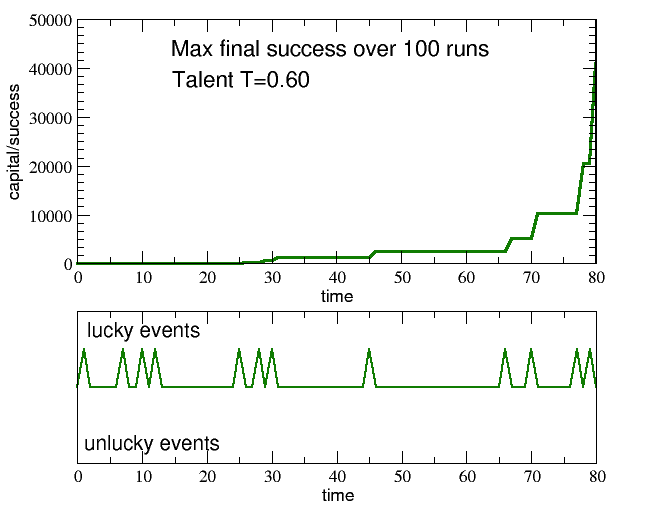}
\caption{\small 
Time evolution of success/capital for the most successful (but moderately gifted) individual over the $100$ simulation runs, compared with the corresponding unusual sequence of lucky events occurred during her working life.
}
\label{global-max-success-evolution} 
\end{center}
\end{figure}

In any case, it is a fact that the absolute best performer over the $100$ simulation runs is an agent with talent $T_{best}=0.6$, perfectly aligned with the average, but with a final success $C_{best}=40960$ which is $650$ times greater than $C_{mt}$ and more than $4000$ times greater than the success $C_{end}<10$ of $2/3$ of the most talented people. This occurs just because, at the end of the story, she was just luckier than the others. Indeed, very lucky, as it is shown in Figure \ref{global-max-success-evolution}, where the increase of her capital/success during her working life is shown, together with the impressive sequence of lucky (and only lucky) events of which, despite the lack of particular talent, she was able to take advantage of during her career. 

Summarizing, what has been found up to now is that, in spite of its simplicity, the TvL model seems able to account for many of the features characterizing, as discussed in the introduction, the largely unequal distribution of richness and success in our society, in evident contrast with the Gaussian distribution of talent among human beings. At the same time, the model shows, in quantitative terms, that a great talent is not sufficient to guarantee a successful career and that, instead, less talented people are very often able to reach the top of success - another "stylised fact" frequently observed real life \cite{Taleb1,Taleb2,Frank}.  

The key point, which intuitively explains how it may happen that moderately gifted individuals achieve (so often) far greater honors and success than much more talented ones, is the hidden and often underestimated role of luck, as resulting from our simulations. But to understand the real meaning of our findings it is important to distinguish the {\it macro} from the {\it micro} point of view. 

In fact, from the {\it micro} point of view, following the dynamical rules of the TvL model, a talented individual has a greater {\it a priori} probability to reach a high level of success than a moderately gifted one, since she has a greater ability to grasp any opportunity will come. Of course, luck has to help her in yielding those opportunities. Therefore, from the point of view of a single individual, we should therefore conclude that, being impossible (by definition) to control the occurrence of lucky events, the best strategy to increase the probability of success (at any talent level) is to broaden the personal activity, the production of ideas, the communication with other people, seeking for diversity and mutual enrichment. In other words, to be an open-minded person, ready to be in contact with others, exposes to the highest probability of lucky events (to be exploited by means of the personal talent).

On the other hand, from the {\it macro} point of view of the entire society, the probability to find moderately gifted individuals at the top levels of success is greater than that of finding there very talented ones, because moderately gifted people are much more numerous and, with the help of luck, have - globally - a statistical advantage to reach a great success, in spite of their lower individual {\it a priori} probability.     

In the next section we will address such a {\it macro} point of view, by exploring the possibilities offered by our model to investigate in detail new and more efficient strategies and policies to improve the average performance of the most talented people in a population, implementing more efficient ways of distributing prizes and resources. In fact, being the most talented individuals the engine of progress and innovation in our society, we expect that any policy able to improve their level of success will have a beneficial effect on the collectivity.

\section{Effective strategies to counterbalance luck}

 The results presented in the previous section are strongly consistent with largely documented empirical evidences, discussed in the introduction, which firmly question the naively meritocratic assumption claiming that the natural differences in talent, skill, competence, intelligence, hard work or determination are the only causes of success. As we have shown, luck also matters and it can play a very important role. The interpretative point is that, being individual qualities difficult to be measured (in many cases hardly defined in rigorous terms), the meritocratic strategies used to assign honors, funds or rewards are often based on individual performances, valued in terms of personal wealth or success. Eventually, such strategies exert a further reinforcing action and pump up the wealth/success of the luckiest individuals through a positive feedback mechanism, which resembles the famous "rich get richer" process (also known as  "Matthew effect" \cite{Merton1,Merton2,Bol}), with an unfair final result.

Let us consider, for instance, a publicly-funded research granting council with a fixed amount of money at its disposal.  In order to increase the average impact of research, is it more effective to give large grants to a few apparently excellent researchers, or small grants to many more apparently ordinary researchers? A recent study \cite{Fortin}, based on the analysis of four indices of scientific impact involving publications, found that impact is positively, but only weakly, related to funding. In particular, impact per dollar was lower for large grant-holders and the impact of researchers who received increases in funding did not  increase in a significant way. The authors of the study conclude that scientific impact (as reflected by publications) is only weakly limited by funding and suggest that funding strategies targeting diversification of ideas, rather than "excellence", are likely to be more productive. A more recent contribution \cite{Mongeon} showed that, both in terms of the quantity of papers produced and of their scientific impact, the concentration of research funding generally produces diminishing marginal returns and also that the most funded researchers do not stand out in terms of output and scientific impact. Actually, such conclusions should not be a surprise in the light of the other recent finding \cite{Sinatra} that impact, as measured by influential publications, is  randomly distributed within a scientist's temporal sequence of publications. In other words, if luck matters, and if it matters more than we are willing to admit, it is not strange that meritocratic strategies reveal less effective than expected, in particular if we try to evaluate merit \textit{ex-post}. In previous studies \cite{Pluchino1,Pluchino2,Pluchino3,Biondo1,Biondo2,Biondo3,Biondo4,Biondo5},  there was  already a warning  against this sort of "naive meritocracy", showing the effectiveness of alternative strategies based on random choices in management, politics and finance. Consistently with such a perspective, the TvL model shows how the minimum level of success of the most talented people can be increased, in a world where luck is important and serendipity is often the cause of important discoveries. 

\subsection{Serendipity, innovation and efficient funding strategies}

The term "serendipity" is commonly used in the literature to refer to the historical evidence that very often researchers make unexpected and beneficial discoveries by chance, while they are looking for something else \cite{Merton,Murayama}. There is a long anecdotal list of discoveries made just by lucky opportunities: from penicillin by Alexander Fleming to radioactivity by Marie Curie, from cosmic microwave background radiation by radio astronomers Arno Penzias and Robert Woodrow Wilson to the graphene by Andre Geim and Kostya Novoselov. Just to give a very recent example, a network of fluid-filled channels in the human body, that may be a previously-unknown organ and that seems to help transport cancer cells around the body, was discovered by chance, from routine endoscopies \cite{Benias}.Therefore, many people think that curiosity-driven research should always be funded, because nobody can really know or predict where it can lead to \cite{Flexner}. 

Is it possible to quantify the role of serendipity? Which are the most efficient ways to stimulate serendipity? 
Serendipity can take on many forms, and it is difficult to constrain and quantify. That is why, so far, academic research has focused on serendipity in science mainly as a philosophical idea. But things are changing. The European Research Council has recently given to the biochemist Ohid Yaqub a $1.7$ million US dollars grant to quantify the role of serendipity in science \cite{Lucky-science}. Yaqub found that it is possible to classify serendipity into four basic types \cite{Yaqub} and that there may be important factors affecting its occurrence. His conclusions seem to agree with ideas developed in earlier works \cite{Page,Cimini,Curry,Nicholson,Bollen,Garner}  which argues that the commonly adopted - apparently meritocratic - strategies, which pursuit excellence and drive out variety, seem destined to be loosing and inefficient. The reason is that they cut out {\it a priori} researches that initially appear less promising but that, thanks also to serendipity, could be extremely innovative {\it a posteriori}. 

From this perspective, we want to use the TvL model, which naturally incorporates luck (and therefore also serendipity) as a quantitative tool for policy, in order to explore, in this subsection, the effectiveness of different funding scenarios. In particular, in contexts where, as above discussed, averagely-talented-but-lucky people are often more successful than more-gifted-but-unlucky individuals, it is important to evaluate the efficiency of funding strategies in preserving a minimum level of success also for the most talented people, who are expected to produce the most progressive and innovative ideas. 

Starting from the same parameters setup used in subsection 2.2, i.e.$N=1000$, $m_T=0.6$, $\sigma_T=0.1$, $I=80$, $\delta_t=6$, $C(0)=10$, $N_E=500$, $p_L=50\%$ and $100$ simulation runs, let us imagine that a given total funding capital $F_T$ is periodically distributed among individuals following different criteria. For example, funds could be assigned:

\begin{enumerate}
\item in equal measure to all  ({\it egalitarian criterion}), in order to foster research diversification; 

\item only to a given percentage of the most successful ("best") individuals ({\it elitarian criterion}), which has been previously referred to "naively" meritocratic, for it distributes funds to people according to their past performance;    

\item by distributing a "premium" to a given percentage of the most successful individuals and the remaining amount in smaller equal parts to all the others  ({\it mixed criterion});

\item only to a given percentage individuals, randomly selected ({\it selective random criterion}); 
\end{enumerate}

 We realistically assume that the total capital $F_T$ will be distributed every $5$ years, during the $40$ years spanned by each simulation run, so that $F_T/8$ units of capital will be allocated from time to time. Thanks to the periodic injection of these funds, we intend to maintain a minimum level of resources for the most talented agents. Therefore, a good indicator, for the effectiveness of the adopted funding strategy, could be the percentage $P_T$, averaged over the $100$ simulation runs, of individuals with talent $T > m_T + \sigma_T$ whose final success/capital is greater than the initial one, i.e. $C_{end}>C(0)$. 
 
 This percentage has already been calculated, in the multiple runs simulation presented in section 2.2. There, we have shown that, in absence of funding, the best performance was scored by very lucky agents with a talent close to the mean, while the capital/success of the most talented people always remained very low. In particular, only a percentage $P_{T0}\sim32\%$ of the total number of agents with $T > 0.7$ reached, at the end of the simulation, a capital/success greater then the initial one. Hence, in order to compare the efficiency of different funding strategies, the increment in the average percentage $P_T$ of talented people which, during their career, increase their initial capital/success should be calculated with respect to $P_{T0}$. Let us define this increment as $P^*_T=P_T-P_{T0}$. The latter quantity is a very robust indicator: we have checked that repeating the set of $100$ simulations, the variation in the value of $P^*_T$ remains  under $2\%$. Finally, if one considers the ratio between $P^*_T$ and the total capital $F_T$ distributed among all the agents during the $40$ years, it is possible to obtain an \textit{efficiency index} $E$, which quantifies the increment of sufficiently successful talented people per unit of invested capital, defined as  $E = P^*_T / F_T$. 

\begin{figure}[t]
\begin{center}
\includegraphics[width=6.00in,angle=0]{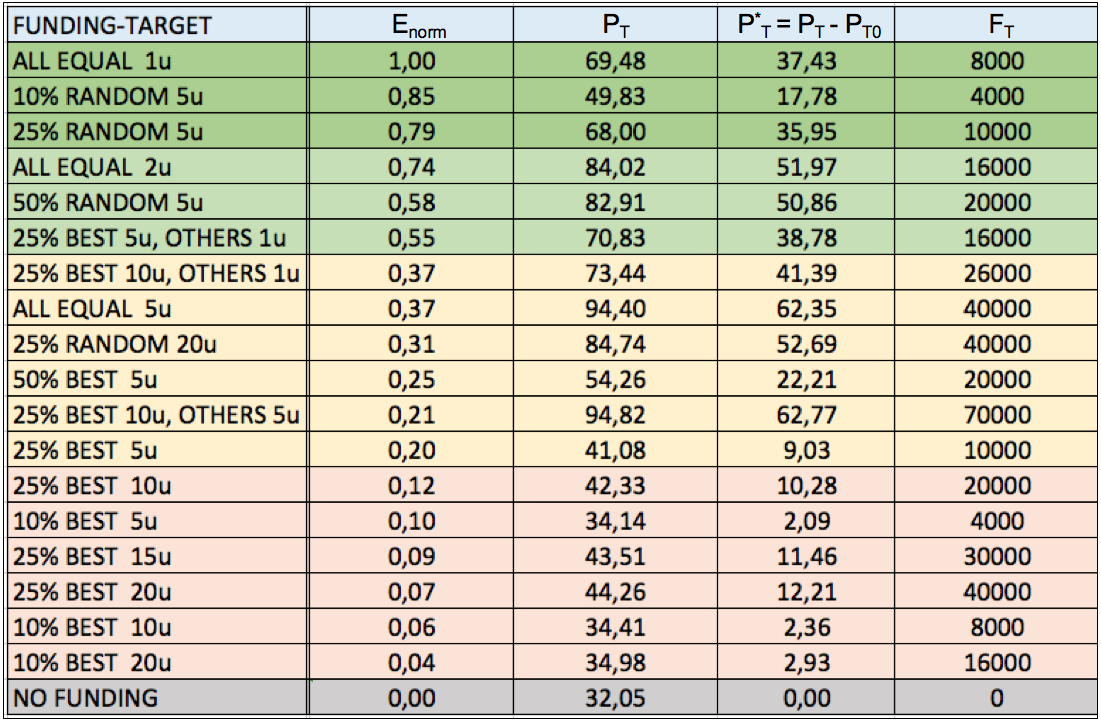}
\caption{\small Funding strategies Table. The outcomes of the normalized efficiency index $E_{norm}$ are reported (2nd column) in decreasing order, from top to bottom, for several funding distribution strategies with different targets (1st column). The corresponding values of both the percentage $P_T$ of successful talented people and its net increase $P^*_T$ with respect to the "no funding" case, averaged over the $100$ simulation runs, are also reported in the third and fourth columns respectively. Finally, the total capital $F_T$ invested in each run, is visible in the last column.      
}
\label{efficiency-table} 
\end{center}
\end{figure}

\begin{figure}[t]
\begin{center}
\includegraphics[width=5.0in,angle=0]{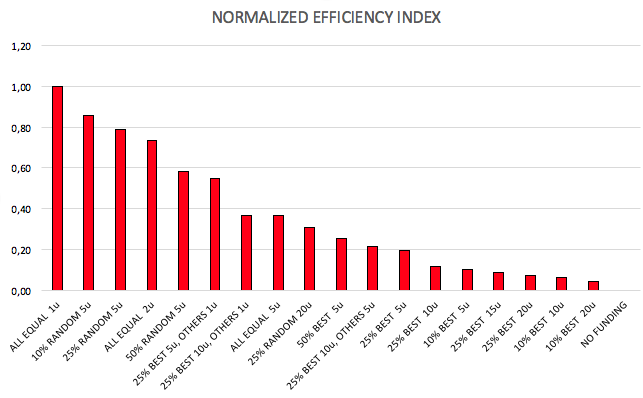}
\caption{\small Normalized Efficiency  index for several funding strategies. The values of the normalized efficiency index $E_{norm}$ are reported as function of the different funding strategies. The figure shows that for increasing  the success of a larger number of talented people with $C_{end}>C(0)$, it is much more efficient to give a small amount of funds to many individuals instead of giving funds in other more selective ways.    
}
\label{efficiency-histogram} 
\end{center}
\end{figure}

In the table shown in Figure \ref{efficiency-table}, we report the efficiency index  (2nd column) obtained for several funding distribution strategies, each one with a different funding target (1st column), together with the corresponding values of $P_T$ (3rd column) and $P^*_T$ (4th column). The total capital $F_T$ invested in each run is also reported in the last column. The efficiency index $E$ has been normalized to its maximum value $E_{max}$ and the various records (rows) have been ordered for decreasing values of $E_{norm}=E/E_{max}$. For the no funding case, by definition, $E_{norm}=0$. The same scores for $E_{norm}$ are also reported in the form of a histogram in Figure \ref{efficiency-histogram}, as a function of the adopted funding strategies. Thanks to the statistical robustness of $P_T$, which shows fluctuations smaller than $2\%$, the results reported for the efficiency index $E_{norm}$ are particularly stable.     

Looking at the table and at the relative histogram of Figure \ref{efficiency-histogram}, it is evident that, if the goal is to reward the most talented persons (thus increasing their final level of success), it is much more convenient to distribute periodically (even small) equal amounts of capital to all individuals rather than to give a greater capital only to a small percentage of them, selected through their level of success - already reached - at the moment of the distribution. 

On one hand, the histogram shows that  the "egalitarian" criterion, which assigns 1 unit of capital every 5 years to all the individuals is the most efficient way to distribute funds, being $E_{norm}=1$ (i.e. $E=E_{max}$): with a relatively small investment $F_T$ of $8000$ units, it is possible to double the percentage of successful talented people with respect to the "no funding" case, bringing it from $P_{T0}=32.05\%$ to $P_T=69.48\%$, with a net increase $P^*_T=37.43\%$. Considering an increase of the total invested capital (for example, setting the egalitarian quotas to 2 or 5 units), this strategy also ensures a further increment in the final percentage of successful talented people $P_T$ (from $69.48\%$ to $84.02\%$ and to $94.40\%$), even if the normalized efficiency progressively decreases from $E_{norm}=1$ to $E_{norm}=0.74$ and to $E_{norm}=0.37$.    

On the other hand, the "elitarian" strategies which assign every 5 years more funds (5, 10, 15 or 20 units) only to the best $50\%$, $25\%$ or even $10\%$ of the already successful individuals, are all at the bottom of the ranking, with $E_{norm}<0.25$: in all of these cases, the net increase $P^*_T$ in the final number of successful talented people with respect to the "no funding" case remains very small (in almost all the cases smaller than $20\%$), often against a much larger invested capital if compared to that of the egalitarian strategy. These results do reinforce the thesis that this kind of approach is only apparently - i.e. naively - meritocratic. 

It is worth noticing that the adoption of a "mixed" criterion, i.e. assigning a "meritocratic" funding share to a certain percentage of the most successful individuals, for instance $25\%$, and distributing the remaining funds in equal measure to the rest of people, gives back better scores for the efficiency index values with respect to the "naively meritocratic" approach. However, the performance of this strategy is not able to overtake the "egalitarian" criterion. As it clearly appears - for example - by the comparison between the sixth and the fourth rows of the funding table, in spite of the same overall investment of $16000$ units, the value of $P_T$ obtained with the mixed criterion stays well below the one obtained with the egalitarian approach ($70.83\%$ against $84.02\%$), as also confirmed by the values of the corresponding efficiency index $E_{norm}$ (0.55 against 0.74). 

If one considers psychological factors (not modeled in this study), a mixed strategy could be revalued with respect to the egalitarian one. Indeed, the premium reward - assigned to the more successful individuals - could induce all agents towards a greater commitment, while the equally distributed part would play a twofold role: at the individual level, it would act in fostering variety and providing unlucky talented people with new chances to express their potential, while feeding serendipity at the aggregate level, thus contributing to the progress of research and of the whole society.             

Looking again at the funding strategy table, it is also worthwhile to stress the surprising high efficiency of the random strategies, which occupy two out of  the three best scores in the general ranking. It results that, for example, a periodic reward of 5 units for only the $10\%$ of randomly selected individuals, with a total investment of just $4000$ units, gives a net increase $P^*_T=17,78\%$, which is greater than almost all those obtained with the elitarian strategies. Furthermore, increasing to $25\%$ the percentage of randomly funded people and doubling the overall investment (bringing it to $10000$ units), the net increase $P^*_T=35.95\%$ becomes comparable to that obtained with the best egalitarian strategy, first in the efficiency ranking. It is striking to notice that this latter score for $P^*_T$ is approximately four times grater than the value ($P^*_T=9.03\%$) obtained with the elitarian approach (see 12th row in the table), distributing exactly the same capital ($10000$ units) to exactly the same number of individuals ($25\%$ of the total). The latter is a further confirmation that, in complex social and economical contexts where chance plays a relevant role, the efficiency of alternative strategies based on random choices can easily overtake that of standard strategies based on the "naively meritocratic" approach. Such a counterintuitive phenomenon, already observed in management, politics and finance (\cite{Pluchino1,Pluchino2,Pluchino3,Biondo1,Biondo2,Biondo3,Biondo4,Biondo5}), finds therefore new evidence also in the research funding context.          

\begin{figure}[t]
\begin{center}
\includegraphics[width=6.00in,angle=0]{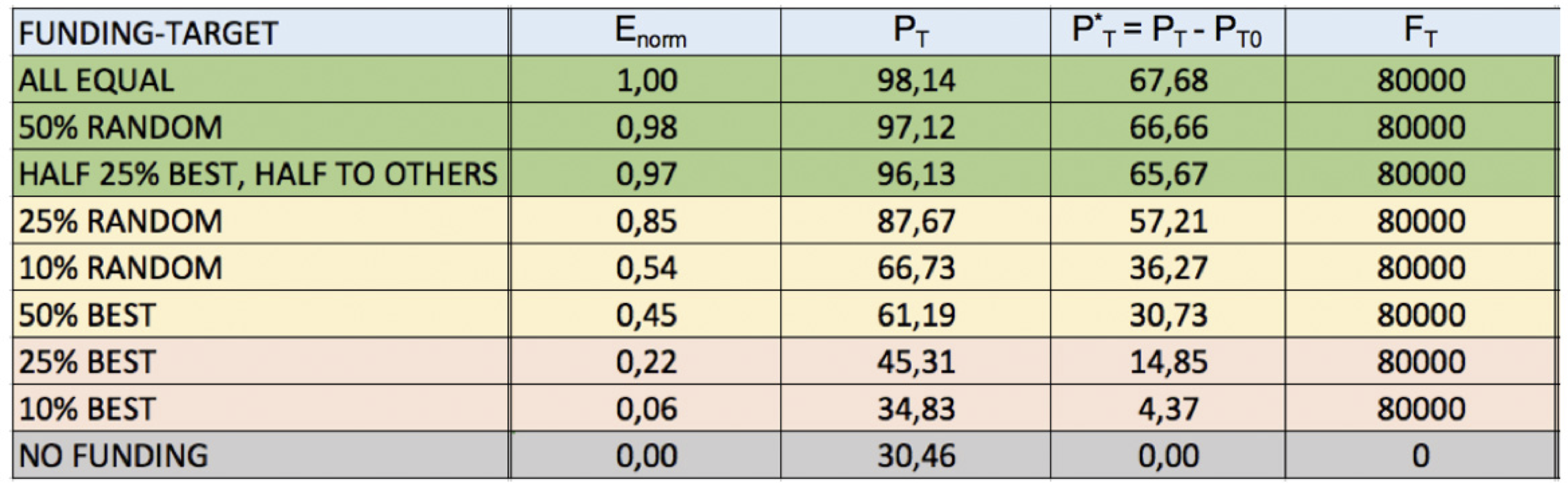}
\caption{\small Funding strategies Table with fixed funds. The outcomes of the normalized efficiency index $E_{norm}$ are reported again in decreasing order, from top to bottom, for several funding distribution strategies with different targets (1st column). At variance with Fig. \ref{efficiency-table}, now the total capital invested in each run was fixed to $F_T=80000$. The egalitarian strategy is, again, at the top of the ranking. 
}
\label{efficiency-table2} 
\end{center}
\end{figure}

To further corroborate these findings, in Figure \ref{efficiency-table2}, the results of another set of simulations are presented. At variance with the previous simulations, the total capital invested in each one of the $100$ runs is now fixed to $F_T=80000$, so that $F_T /8 = 10000$ units are distributed every 5 years among the agents following the main funding strategies already considered. Looking at the table, the egalitarian strategy results again the most efficient in rewarding the most talented people, with a percentage $P_T$ close to $100\%$, immediately followed by the random strategy (with $50\%$ of randomly funded individuals) and by the mixed one, with half of the capital distributed to the $25\%$ of the most successful individuals and the other half in equal measure to the remaining people. On the contrary, all the elitarian strategies are placed again at the bottom of the ranking, thus further confirming the inefficiency of the "naively meritocratic" approach in rewarding real talent.

The results of the TvL model simulations presented in this subsection, have focused on the importance of external factors (as, indeed, efficient funding policies) in increasing the opportunities of success for the most talented individuals, too often penalized by unlucky events. In the next subsection we investigate to what extent new opportunities can be originated by changes in the environment as for example the level of education or other stimuli received by the social context where people live or come from.     

\subsection{The importance of the environment}

First, let us estimate the role of the average level of education among the population. Within the TvL model, the  latter  could be obtained  by changing  the parameters of the normal distribution of talent. Actually, assuming that talent and skills of individuals, if stimulated, could be more effective in exploiting new opportunities, an increase in either the mean $m_T$ or the standard deviation $\sigma_T$ of the talent distribution could be interpreted as the effect of policies targeted, respectively, either at raising the average level of education or at reinforcing the training of the most gifted people.

\begin{figure}[t]
\begin{center}
\includegraphics[width=4.5in,angle=0]{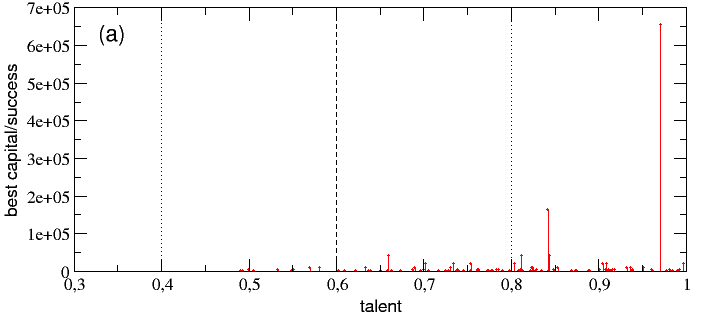}
\includegraphics[width=4.6in,angle=0]{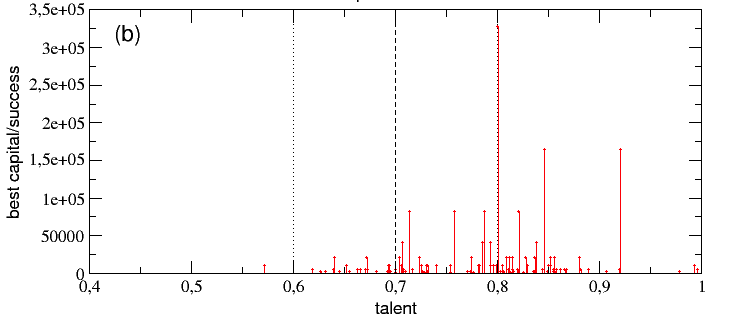}
\caption{\small 
The final capital of the most successful individuals in each of the $100$ runs is reported as function of their talent for populations with different talent distributions parameters: (a) $m_T=0.6$ and $\sigma_T=0.2$ (which represent a training reinforcement for the most gifted people); (b) $m_T=0.7$ and $\sigma_T=0.1$ (which represents an increase in the average level of education). The corresponding $m_T$ and $m_T \pm \sigma_T$ values are also indicated as, respectively, vertical dashed and dot lines. 
}
\label{education-degree} 
\end{center}
\end{figure}

In the two panels of Figure \ref{education-degree} we report the final capital/success accumulated by the best performers in each of the $100$ runs, as function of their talent. The parameters setup is the same than in subsection 2.2 ($N=1000$, $I=80$, $\delta_t=6$, $C(0)=10$, $N_E=500$ and $p_L=50\%$) but with different moments for the talent distributions. In particular, in panel (a) we left unchanged $m_T=0.6$ but  increased $\sigma_T=0.2$, while in panel (b) we made the opposite, leaving $\sigma_T=0.1$ but  increasing $m_T=0.7$. In both cases, a shift on the right of the maximum success peaks can be appreciated, but with different details. 

Actually, it results that increasing $\sigma_T$ without changing $m_T$, as shown in panel (a), enhances the chances for more talented people to get a very high success: the best performer is, now, a very talented agent with $T=0.97$, who reaches an incredible level of capital/success $C_{best}=655360$. This, on one hand, could be considered positive but, on the other hand, it is an isolated case and it has, as a counterpart, an increase in the gap between unsuccessful and successful people.

Looking now at panel (b), it results that increasing $m_T$ without changing $\sigma_T$ produces a best performer, with $C_{best}=327680$ and a talent $T =0.8$, followed by other two with $C=163840$ and, respectively, $T=0.85$ and $T=0.92$. This means that also in this case the chances for more talented people to get a very high success are enhanced, while the gap between unsuccessful and successful people is lower than before.

 Finally, in both considered examples, the average value of the capital/success for the most talented people over the $100$ runs is increased with respect to the value $C_{mt}\sim63$ found in subsection 2.2. In particular, we found $C_{mt}\sim319$ for panel (a) and $C_{mt}\sim122$ for panel (b), but these values are quite sensitive to the specific set of simulation runs. A more reliable parameter in order to quantify the effectiveness of the social policies investigated here is, again, the indicator $P_T$ introduced in the previous subsection, i.e. the average percentage of individuals with talent $T > m_T + \sigma_T$ and with final success/capital $C_{end}>10$, over the total number of individuals with talent $T > m_T + \sigma_T$ (notice that now, in both the cases considered, $m_T + \sigma_T = 0.8$). In particular, we found $P_T=38\%$ for panel (a) and $P_T=37.5\%$ for panel (b), with a slight net increment with respect to the reference value $P_{T0}=32\%$ (obtained for a talent distribution with $m_T=0.6$ and $\sigma_T=0.1$).    
 
\begin{figure}[t]
\begin{center}
\includegraphics[width=4.6in,angle=0]{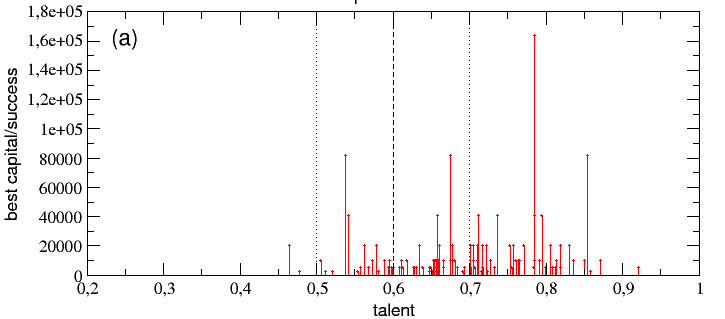}
\includegraphics[width=4.55in,angle=0]{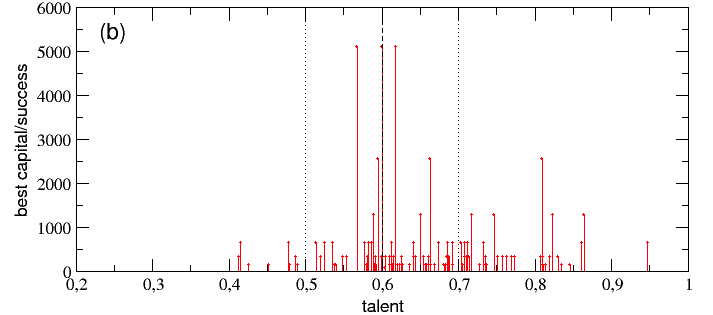}
\caption{\small 
The final capital of the most successful individuals in each of the $100$ runs is reported as function of their talent, for populations living in environments with a different percentage $p_L$ of lucky events: (a) $p_L=80\%$; (b) $p_L=20\%$. The values of $m_T=0.6$ and $m_T \pm \sigma_T$, with $\sigma_T=0.1$ are also indicated as, respectively, vertical dashed and dot lines.
}
\label{country-of-birth} 
\end{center}
\end{figure}
 
Summarizing, our results indicate that  strengthening the training of the most gifted people or  increasing the average level of education  produce, as one could expect, some beneficial effects on the social system, since both these policies raise the probability, for talented individuals, to grasp the opportunities that luck presents to them. On the other hand, the enhancement in the average percentage of highly talented people who are able to reach a good level of success, seems to be not particularly remarkable in both the cases analyzed, therefore the result of the corresponding educational policies appears mainly restricted to the emergence of isolated extreme successful cases. 
 
 Of course, once a given level of education has been fixed, it is quite obvious that the abundance of opportunities offered by the social environment, i.e. by the country where someone accidentally is born or where someone choose to live, it is another key ingredient able to influence the global performance of the system. 
    
In Figure \ref{country-of-birth} we show results analogous to those shown in the previous figure, but for another set of simulations, with $100$ runs each, with the same parameters setup as in subsection 2.2 ($N=1000$, $m_T=0.6$, $\sigma_T=0.1$, $I=80$, $C(0)=10$, $N_E=500$) and with different percentages $p_L$ of lucky events (we remind that, in subsection 2.2., this percentage was set to $p_L=50\%$). In panels (a) we set $p_L=80\%$, in order to simulate a very stimulating environment, rich of opportunities, like that of rich and industrialized countries such as the U.S. \cite{Milanovic}. On the other hand, in panels (b), the value $p_L=20\%$ reproduces the case of a much less stimulating environment, with very few opportunities, like for instance that of Third World countries. 

As visible in both panels, the final success/capital of the most successful individuals as function of their talent strongly depend on $p_L$. 

When $p_L=80\%$, as in panel (a), several agents with medium-high talent are able to reach higher levels of success compared to the case $p_L=50\%$, with a peak of $C_{best}=163840$. On the other hand, the average value of the capital/success for the most talented individuals, $C_{mt}\sim149$, is quite high and, what is more important, the same holds for the indicator $P_T=62.18\%$ (about twice with respect to the reference value $P_{T0}=32\%$), meaning that, as expected, talented people benefits of the higher percentage of lucky events. 

Completely different outcomes are obtained with $p_L=20\%$. Indeed, as visible in panel (b), the overall level of success is now very low, if compared to that found in the simulations of subsection 2.2, with a peak value $C_{best}$ of only 5120 units: it is a footprint of a reduction in the social inequalities, which is an expected consequence of the flattening of success opportunities. According with these results, also the $P_T$ indicator reaches a minimal value, with an average percentage of only $8.75\%$ of talented individuals able to increase their initial level of success.    

In conclusion, in this section we have shown that a stimulating environment, rich of opportunities, associated to an appropriate strategy for the distribution of funds and resources, are important factors in exploiting the potential of the most talented people, giving them more chances of success with respect to the moderately gifted, but luckier, ones. At the macro level, any policy able to influence those factors and to sustain talented individuals, will have the result of ensuring collective progress and innovation.

\section{Conclusive remarks}

In this paper, starting from few very simple and reasonable assumptions, we have presented an agent-based model which is able to quantify the role of talent and luck in the success of people's careers. The simulations show that although talent has a Gaussian  distribution among agents,  the resulting  distribution of success/capital after a working life of 40 years, follows a power law which respects the "80-20" Pareto law for the distribution of wealth found in the real world. An important result of the simulations is that the most successful agents are almost never the most talented ones, but those around the average of the Gaussian talent distribution - another stylised fact often reported in the literature. The model shows the importance, very frequently underestimated, of lucky events in determining the final level of  individual success. Since rewards and resources are usually given to those that have already reached a high level of success, mistakenly considered as a measure of competence/talent, this result is even a more harmful disincentive, causing a lack of opportunities for the most talented ones. Our results highlight the risks of the paradigm that we call  "naive meritocracy", which  fails to give honors and rewards to the most competent people, because it underestimates the role of randomness among the determinants of success. In this respect, several different scenarios have been investigated in order to discuss more efficient strategies, which are able  to counterbalance the unpredictable role of luck and give more opportunities and resources to the most talented ones - a purpose that should be the main aim of a truly meritocratic approach. Such strategies have also been shown to be the most beneficial for the entire society, since they tend to increase the diversity of ideas and perspectives  in research, thus fostering also innovation.

\section*{Acknowlegments}
We would like to thank Robert H. Frank, Pawel Sobkowicz and Constantino Tsallis for fruitful discussions and comments.


\end{document}